\documentclass[prd,nofootinbib,a4paper,showpacs,showkeys,twocolumn]{revtex4}
\usepackage{hyperref}

\usepackage{graphicx}
\usepackage{amsmath}
\usepackage{amssymb}
\usepackage{amsfonts}
\usepackage{mathrsfs}
\usepackage{mathtools}

\usepackage{units}
\usepackage[dvipsnames]{xcolor}

\newcommand{\Mpl}{M_\mathrm{Pl}}
\newcommand{\eqn}{Eq.\,}
\newcommand{\CP}{\textit{CP}}

\begin{document}



\title{Leptogenesis in models with keV sterile neutrino dark matter}

\author{F. Bezrukov$^{a,b,c}$}
\email[\,]{fedor.bezrukov@uconn.edu}

\affiliation{%
$^a$Physics Department, University of Connecticut, Storrs, CT
06269-3046, USA\\
$^b$RIKEN-BNL Research Center, Brookhaven National Laboratory, Upton,
NY 11973, USA\\
$^c$Ludwig-Maximilians-Universit\"at, Theresienstra\ss{}e 37, 
80333, M\"unchen, Germany}

\author{A. Kartavtsev$^{d}$}
\email[\,]{alexander.kartavtsev@mpi-hd.mpg.de}

\author{M. Lindner$^{d}$}
\email[\,]{manfred.lindner@mpi-hd.mpg.de}

\affiliation{%
$^d$Max-Planck-Institut f\"ur Kernphysik, Saupfercheckweg 1, 69117 Heidelberg,
Germany}

\begin{abstract}
We analyze leptogenesis in gauge extensions of the Standard Model 
with keV sterile neutrino dark matter. We find that both the observed 
dark matter abundance and the correct baryon asymmetry of the Universe
can simultaneously emerge in these models.
Both the dark matter abundance and the leptogenesis are controlled by
the out of equilibrium decays of the same heavy right handed neutrino. 
\end{abstract}

\maketitle


\section{Introduction}

The modern cosmology provides at least two experimental evidences for the existence 
of new physics beyond the Standard Model (SM).  One of them is  the baryon asymmetry 
of the Universe (BAU). The other one is the Dark Matter (DM) which constitutes most 
of the gravitating mass in the Universe.  To be consistent with the observations, the 
DM candidate should be a very weakly interacting particle. For a sterile neutrino  
DM-candidate there are two ways to achieve the observed DM abundance. The first 
possibility is that the sterile neutrino interacts so weakly, that it never enters 
thermal equilibrium after the reheating ($\nu$MSM situation
\cite{Asaka:2005an,Asaka:2005pn,Boyarsky:2009ix,Bezrukov:2009yw,Shaposhnikov:2006xi,Canetti:2012vf} 
or models with extra-dimensions \cite{Kusenko:2010ik}, see also section \textbf{1~D} of  
\cite{SNAC} for a more comprehensive review).
Here we are going to pursue another interesting  possibility, described in
\cite{Bezrukov:2009th}.
We assume in this paper that the
SM gauge group is embedded in some larger group (for concreteness 
we consider a left-right symmetric group $SU(2)_L\times
SU(2)_R\times U(1)$) and the right handed neutrinos are charged under this
group:
\begin{align}
  \label{eq:Lcc}
  -{\cal L}_{CC} = \frac{g}{\sqrt{2}} \sum_a \left(
    W_L^\mu \,\overline{l_{aL}}\gamma_\mu\nu_{aL}
    +W_R^\mu \,\overline{l_{aR}}\gamma_\mu N_{aR}
  \right)\nonumber\\
  +\textrm{h.c.}  
\end{align}
This additional gauge interaction brings all the right handed
neutrinos into the thermal equilibrium at some high temperature.
In such a scenario the right-handed neutrinos are expected
to be heavy and the active-sterile mixings, which are important 
for our scenario, are generally expected to be tiny. 
If one of the sterile neutrinos (called $N_1$) is light (keV scale),
it can be stable enough to be a DM candidate by a reasonable tuning 
of the coupling constants.  However, being a light thermal relic, 
one may expect that it is significantly overproduced.  This problem 
is, however, not an issue if there is a later dilution of the 
overproduced amount of the DM neutrino by a subsequent 
out-of-equilibrium decay of another massive particle.  A good
and most natural candidate for such a particle is a heavier right handed
neutrino ($N_2$).  If it also has a sufficiently small mixing angle
with left handed neutrinos then it freezes out at high temperatures
together with $N_1$ (due to identical gauge interactions).
Later it decays to the SM particles while being significantly out of equilibrium,
thus diluting the abundance of the DM sterile neutrino $N_1$.
Models of this type are able to reproduce the experimentally 
observed DM abundance provided that certain constraints on the model parameters are 
imposed \cite{Bezrukov:2009th}. In short,  there are very strict bounds on the Yukawa 
couplings of the DM sterile neutrino which are required to suppress its radiative decay
constrained by the X-ray observations (see \cite{Boyarsky:2009ix} for a review and
\cite{Boyarsky:2006fg,Boyarsky:2006ag,Boyarsky:2007ay,Boyarsky:2006hr,Boyarsky:2007ge,Watson:2006qb,Yuksel:2007xh} for details) and bounds on the Yukawa  couplings of the other sterile neutrino appearing 
from the requirement of the entropy generation (which controls the DM density).
In addition there are bounds on the masses of the sterile neutrinos and additional
gauge bosons, appearing from the fact that the entropy generation should happen before 
the Big Bang Nucleosynthesis (see section 2.7 of \cite{Bezrukov:2009th}).

In addition to providing  a DM candidate, any viable extension of the SM should also be 
able to explain the observed BAU. In models with right-handed sterile neutrinos it is 
rather natural to expect additional \CP-violation in the neutrino sector 
leading to the generation of a lepton asymmetry (leptogenesis), which later 
on is transferred to the
baryon asymmetry by the sphaleron processes.  The simplest possibility here
is the usual thermal leptogenesis at high temperatures, associated with
\CP-asymmetric decays of the heavy sterile neutrinos.  
The main goal of this paper is to demonstrate
that the constraints imposed by the requirement of successful 
dark matter production are compatible with leptogenesis.
Thus both the dark matter of the Universe and the BAU can be explained together 
in a rather minimalistic scenario where heavy sterile neutrinos decay to one 
light sterile neutrino. 
We also find additional constraints on the
neutrino masses, appearing from the requirement of successful 
leptogenesis and
analyze possible patterns of masses and decay widths
of the sterile neutrinos in the model.

\section{Constraints from leptogenesis}
\label{sec:constr-from-lept}
The lepton asymmetry generated in the decay 
of the heavy neutrino \cite{Sakharov:1967dj,Fukugita:1986hr} is converted into 
the BAU by sphaleron processes \cite{Kuzmin:1985mm}.
This implies that the generation of the lepton
asymmetry should happen before sphaleron freezeout
($T\simeq\unit[100]{GeV}$), and that the sterile neutrinos should be out of
thermal equilibrium at higher temperatures.  These two requirements
are similar to the BBN constraints found in \cite{Bezrukov:2009th}, where it was
required that the heavy sterile neutrinos decay before the BBN.
In complete analogy with \cite{Bezrukov:2009th} we require that
sterile neutrino $N_{2}$ decays into the SM particles
before sphaleron freezeout and 
generates sufficient entropy to dilute the DM sterile neutrino ($N_1$)
abundance. Replacing the BBN temperature by the sphaleron freezeout temperature
(denoted by $T_r$ here) in Eqs.~(20--22) of \cite{Bezrukov:2009th} we find:
\begin{equation}
  \label{eq:2}
  M_2 > 200\left(\frac{M_1}{\unit[1]{keV}}\right)T_r
      \sim \unit[2\times10^4]{GeV}
  \;.
\end{equation}
In addition, the requirement of efficient entropy production 
in the decays of $N_2$ implies that it should freeze out while being still 
relativistic, $T_\mathrm{f}>M_2$.
This constraints the strength of the gauge interaction of the right handed neutrinos or,
equivalently, the mass $M_R$ of the additional gauge bosons. Although
this requirement is similar to those in
\cite{Bezrukov:2009th}, the lower 
bound on the neutrino mass $M_2$ is now larger 
and the lowest possible value for $M_R$ is higher
\begin{equation}
  \label{eq:3}
  \frac{M_R}{g_R} > \frac{1}{g_{*f}^{1/8}}\left(\frac{M_2}{\unit[2\times10^4]{GeV}}\right)^{3/4}
      \unit[10^4]{TeV}
  \;,
\end{equation}
where $g_R$ is the gauge coupling constant for the additional gauge group.
All other constraints formulated in \cite{Bezrukov:2009th} for
successful DM generation are left intact.

To achieve proper entropy dilution, the entropy generating neutrino $N_2$
should decouple while relativistic and has decay width
\begin{equation}
\label{Gamma2}
  \Gamma_2 \simeq
  0.50\times10^{-6}
  \frac{g_{*\mathrm{f}}^2}{g_{*}^2}
  \bar{g}_{*}^{1/2}\frac{M_2^2}{\Mpl}
  \left(\frac{1\unit{keV}}{M_1}\right)^2
  \;,
\end{equation}
which can be used to calculate the Yukawa couplings of  $N_2$.

\section{\label{sec:estim-lept}Lepton asymmetry}
In Friedman-Robertson-Walker Universe the evolution equation for the particle number 
density in the comoving volume, $Y\equiv \sqrt{-g_3}\,n$, has the form \cite{Kartavtsev:2008fp}
\begin{align}
\frac{dY}{dz}=\frac{\sqrt{-g_3}}{\dot z}\int \frac{\hat C[f]}{E} d\Omega_p\,,
\end{align}
where $\sqrt{-g_3}=a^3\propto s^{-1}$ is the determinant of the spatial part of the metric,
$s$ is the entropy density, $z=M/T$ is the dimensionless inverse temperature, 
$\dot z$ 
is its derivative with respect to the proper time $\tau$ and finally $\hat C$ 
the collision term. 
Under the usual assumption that the distribution function of the Majorana 
neutrino is proportional to the equilibrium one, the contribution of the decay 
and inverse decay processes to the lepton asymmetry reads \cite{Kartavtsev:2008fp}: 
\begin{align}
\label{collisiontermLept}
\int \frac{\hat C[f_L]}{E} d\Omega_p\,=\sum_i\langle \Gamma_{N_i}\rangle \bigl[
\epsilon_i (n_{N_i}&-n_{N_i}^{eq})\nonumber\\
&-n_L\, n_{N_i}^{eq}/n_{L}^{eq}\bigr]\,,
\end{align}
where $\epsilon$ is the  \CP-violating parameter and $\langle \Gamma_{N_i}\rangle$
is the thermally averaged decay width of the heavy neutrino. The first term in the 
square brackets describes the generation of the lepton asymmetry and is proportional  
to the deviation of the Majorana neutrinos from thermal equilibrium. The second term describes 
washout effects due to the inverse decays of the heavy neutrinos and is proportional 
to the generated lepton asymmetry $n_L$. 
The contribution of the same processes to the 
right-hand side of the rate equations for the Majorana number density is given by 
\cite{Kartavtsev:2008fp}: 
\begin{align}
\label{collisiontermMaj}
\int \frac{\hat C[f_{N_i}]}{E} d\Omega_p\,=-\langle \Gamma_{N_i}\rangle (n_{N_i}-n_{N_i}^{eq})\,,
\end{align}

In the model under consideration the right-handed neutrinos are
efficiently equilibrated by the $SU(2)_R$ gauge bosons. Before the
freezeout of the gauge interactions the right-handed neutrinos are in
thermal equilibrium and no asymmetry is generated.  After the
freeze-out the requirement that one of the Majorana neutrinos is
dark matter together with the universality of the $SU(2)_R$
gauge interactions implies that all three heavy species are completely
out of equilibrium\footnote{Note that strictly speaking the 
strong Yukawa interactions could alter the freezeout temperature of the 
sterile neutrinos. For the dark matter and the entropy generating 
sterile neutrinos the Yukawas are constrained by the upper bound on their 
lifetimes. The Yukawa couplings of the third sterile neutrino can in 
principle be large, which would lead to a decrease of the generated asymmetry.}
\cite{Bezrukov:2009th}. 
In this case the washout processes play no role and the second term on
the right-hand side of \eqn\eqref{collisiontermLept} can be
neglected. Therefore, the rate equation for the lepton asymmetry takes
a simple form:
\begin{align}
\frac{dY_L}{dz}=-\sum_i\epsilon_i\,\frac{dY_{N_i}}{dz}\,.
\end{align}
Since the initial lepton asymmetry and the final  number densities of $N_{2,3}$ are zero,
its solution reads: 
\begin{align}
  \label{eq:Yfin}
Y^{fin}_L=\sum_i\epsilon_i\,Y^{in}_{N_i}\,.
\end{align}
To convert the particle yield \eqref{eq:Yfin} into the leptonic asymmetries
we should take into account that after the freezeout at
$T_f$ additional entropy was generated by the decays of the heavy neutrinos.
Using \eqref{eq:Yfin} we get for the lepton number density to the
entropy ratio after the decay of $N_{2,3}$
\begin{equation}
  \label{DeltaL}
  \Delta_L \equiv \frac{n_L}{s} = \frac{\epsilon_i
    Y^{in}_{N_i}}{s_fa_f^3}\frac{s_f a_f^3}{s_ea_e^3} =
  \frac{\epsilon_i}{S}\frac{n_{N_i,f}}{s_f}  \;,
\end{equation}
where $n_L$ and $n_N$ are lepton and sterile
neutrino number densities, index $f$ corresponds to the freezeout and
index $e$ to some moment after the heavy neutrino 
has decayed.
Assuming that the sterile neutrino decouples while still being
  relativistic its initial abundance is 
\begin{equation}
\frac{n_{N_i,f}}{s_f}=
  \frac{1}{g_{*}}\frac{135\zeta(3)}{4\pi^4}
  \;.
  \label{eqn:r}
\end{equation}
The entropy generation factor $S$ is due to the out of equilibrium decay of heavy
sterile neutrino $N_i$ \cite{Scherrer:1984fd}
\begin{align}
\label{SDecayWidth}
  S  \simeq   0.76\frac{\bar{g}_{*}^{1/4}M_i}{g_{*}\sqrt{\Gamma_i \Mpl}}
  \;.
\end{align}
Here we would like to stress two points.  First, the formula
\eqref{SDecayWidth} is derived for $S\gg1$, which is motivated 
by the fact that
$S\sim10-100$ is required to obtain a proper DM density \cite{Bezrukov:2009th} 
\begin{equation}
  \label{Sreq}
  S \simeq 100 \left(\frac{10.75}{g_{*\mathrm{f}}}\right)
  \left(\frac{M_1}{\unit[1]{keV}}\right)
  \;.
\end{equation}
Second, we have in principle two entropy generating neutrinos.  It is
easy to check that if the decay rates $\Gamma_2$ and $\Gamma_3$ are
significantly different (while the masses are similar), then the resulting entropy generation is
dominated by the neutrino with the smallest decay width, see \eqref{SDecayWidth}.  
If both decay widths or entropy generation factors $S_1$ and $S_2$ are 
of the same order  the result can be obtained by numerical solution of
the differential equations from \cite{Scherrer:1984fd},
generalized to the multi-species case.

The generated lepton asymmetry is transferred to the baryon asymmetry by sphalerons 
with the coefficient $\Delta B=-0.54\,\Delta L$ \cite{Kuzmin:1985mm,Kuzmin:1987wn,Khlebnikov:1996vj,Dreiner:1992vm,Kartavtsev:2005rs}.
Taking $S\approx {\cal O}(10)$
we then find 
\begin{align}
\Delta B\approx -1.5 \times 10^{-4}\,\epsilon\,,
\end{align}
which is to be compared with the experimentally measured value $\Delta B\approx 0.86\times 10^{-10}$. 
Since the washout processes are strongly suppressed in the considered model, successful
leptogenesis is possible if $\epsilon\approx -6 \times 10^{-7}$.

The \CP-violating parameter in the decay of  the $i$'th neutrino receives two 
contributions. The self-energy contribution is given by \cite{Pilaftsis:2003gt,Frossard:2012pc}:
\begin{equation}
  \label{epsilon_s}
  \epsilon^S_i =-\eta_{ij} \frac{R_{ij}}{R^2_{ij}+A^2_{ij}}\,,\quad 
  R_{ij}\equiv \frac{\Delta M_{ji}^2}{M_i \Gamma_j}\;,
\end{equation}
where $\eta_{ij}$ is defined by  ${\rm Im}(h^\dagger h)^2_{ij}\equiv \eta_{ij}
(h^\dagger h)_{ii}(h^\dagger h)_{jj}$. For the on-shell quasiparticle approximation
that we use here the `regulator' $A$ is given by $A_{ij}=M_i/M_j$, see 
\cite{Garny:2011hg,Frossard:2012pc} for more details. The vertex contribution to 
the \CP-violating parameter reads \cite{Fukugita:1986hr,Frossard:2012pc}:
\begin{align}
\label{epsilon_v}
 \epsilon_i^V=\eta_{ij} \frac{\Gamma_j}{M_j}\,f\biggl(\frac{M_j^2}{M_i^2}\biggr)\,,
\end{align}
where
\begin{align}
f(x)=\sqrt{x}\left[1-(1+x)\ln\left(\frac{1+x}{x}\right)\right]\xrightarrow{x\gg1} \frac{-1}{2\sqrt{x}}\,.
\end{align}

Since the DM candidate has a mass of order of a few keV, the corresponding \CP-violating 
parameter is strongly suppressed by the  mass ratio and its contribution to the lepton asymmetry 
can be neglected. Furthermore, the effective in-medium masses of the Higgs and 
leptons are of order of one tenth of the temperature, so that the decays of the lightest Majorana 
neutrino are kinematically forbidden. Therefore only the two heavier neutrinos $N_{2,3}$ are 
relevant for the asymmetry generation. 

\subsection{Hierarchical mass spectrum}
In the considered model the washout processes are strongly suppressed. Therefore the 
usual argument that the asymmetry generated by the heavier Majorana is washed out 
by the inverse decay and scattering processes mediated by the lighter one is not 
applicable. This means that even for a hierarchical mass spectrum we have to consider 
contributions of both Majorana neutrinos. For a hierarchical mass spectrum the sum of
the self-energy and vertex contributions can be approximated by: 
\begin{align}
\label{EpsilonApprox}
\epsilon_i&\approx -\frac32 \eta_{ij}\frac{M_i}{M_j}\frac{\Gamma_j}{M_j}\quad {\rm if}\quad M_i\ll M_j\,,\\
\epsilon_i&\approx \eta_{ij}\frac{M_j}{M_i}\frac{\Gamma_j}{M_j}\left[2+\ln M_j^2/M_i^2\right]\quad 
{\rm if} \quad M_i\gg M_j\,.
\end{align}
For $M_i \gg M_j$ the $\ln M_j^2/M_i^2$ term is a large negative number and therefore 
both expressions have the same overall sign. In the following we enumerate the heavy 
neutrinos such, that the leading contribution to the entropy generation is due to 
$N_2$, i.e.\  $M_2/\sqrt{\Gamma_2}>M_3/\sqrt{\Gamma_3}$. Then the requirement of 
sufficient entropy dilution, see \eqn\eqref{Gamma2}, implies that 
\begin{align}
\frac{\Gamma_2}{M_2}\sim 10^{-5} \frac{M_2}{M_{Pl}}\,.
\end{align}
On the other hand the ratio 
\begin{align}
\frac{\Gamma_3}{M_3}\equiv\frac{\tilde m_{33}M_3}{8\pi v^2}\,, 
\end{align}
where $\tilde m_{33}=(hh)_{33}\,v^2/M_3$ is the see-saw contribution to the effective 
mass of the active neutrino, is essentially unconstrained. Combined with \eqref{EpsilonApprox} 
this implies that for reasonable masses of the right-handed neutrinos $\epsilon_3$ is always 
strongly suppressed as compared to $\epsilon_2$. In other words, most of the lepton asymmetry 
is generated by $N_2$. Depending on the choice of the model parameters there are three 
distinct situations: 
\begin{enumerate}
\item[a)] If  $\Gamma_2\gg\Gamma_3$ then $M_2\gg M_3$. Assuming maximal \CP-violating phase, 
i.e.\ assuming that $\eta_{ij}\sim 1$,we find for the \CP-violating parameter: 
\begin{align*}
\epsilon_2&\sim \frac{\Gamma_3}{M_2}\left[2+\ln M_3^2/M_2^2\right] \nonumber\\
&\ll  \frac{\Gamma_2}{M_2}\left[2+\ln M_3^2/M_2^2\right] \sim 10^{-5}\frac{M_2}{M_{Pl}}\,.
\end{align*}
Therefore  successful leptogenesis is only possible if $M_2\sim M_{Pl}$ and this case 
is excluded. 

\item[b)] If $\Gamma_3\sim\Gamma_2$ then $M_2>M_3$. The resulting expression for the 
\CP-violating parameter is the same as in the previous case and successful leptogenesis 
is again possible only if $M_2\sim M_{Pl}$.  
  
\item[c)] If $\Gamma_3\gg\Gamma_2$ then both $M_2>M_3$ and $M_2<M_3$ are possible. In the 
former case the \CP-violating parameter is given by:
\begin{align}
\epsilon_2&\sim\frac{\Gamma_3}{M_2} \left[2+\ln M_3^2/M_2^2\right] \nonumber\\
&=\frac{\tilde m_{33}M_3}{8\pi v^2}\frac{M_3}{M_2}\left[2+\ln M_3^2/M_2^2\right]\,,
\end{align}
whereas in the latter case we obtain 
\begin{align*}
\epsilon_2\sim \frac{M_2}{M_3} \frac{\Gamma_3}{M_3}\equiv\frac{\tilde m_{33}M_3}{8\pi v^2}\frac{M_2}{M_3}\;.
\end{align*}
Since the see-saw contributions of the first and second heavy
neutrinos should be small, 
see \cite{Bezrukov:2009th}, $\tilde{m}_{33}$ should not contribute significantly 
to the neutrino masses, i.e.\ should not exceed the atmospheric mass 
difference\footnote{Let us remind the reader that in this model the observed oscillation 
pattern can not be achieved with the type I see-saw and should follow form some 
other mechanism, eg.\, a type II see-saw as anyway expected in left-right symmetric models, 
see \cite{Bezrukov:2009th}.
Then, if $\tilde{m}_{33}$ exceeds the
atmospheric mass difference,
nontrivial cancellations between the type I and type II contributions are required.}. We can see 
that if the hierarchy between $M_2$ and $M_3$ is mild, the 
required asymmetry is produced for $M_{2,3}\sim\unit[10^9]{GeV}$, which is well 
below the GUT scale. That is, in this case the \CP-violating parameter can be large 
enough to generate the required amount of the baryon asymmetry. Note also that the 
gauge interaction scale bound \eqref{eq:3}, corresponding to this value, is also 
below the GUT scale, $M_R\gtrsim\unit[10^{10}]{GeV}$.
\end{enumerate}
From the above analysis it follows that for a hierarchical mass spectrum 
\begin{figure}[h!]
\includegraphics[width=0.95\columnwidth]{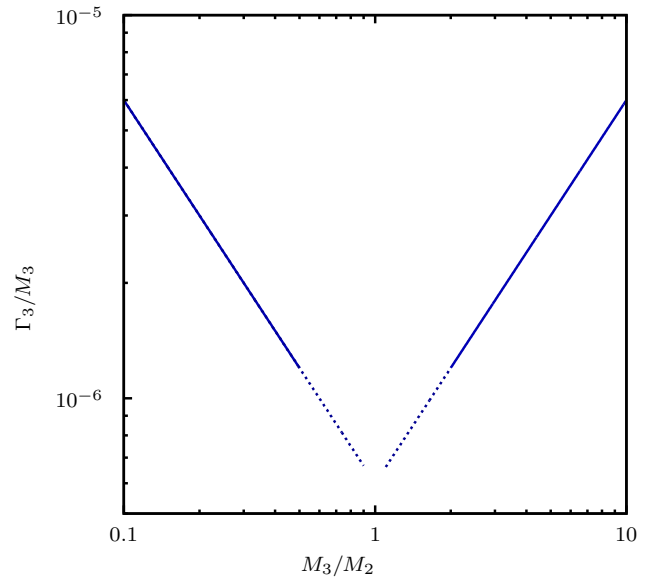}
\caption{\label{ParameterSpace} Schematic plot of the parameter space 
where correct baryon asymmetry and dark matter abundance are realized.
The dots correspond to quasidegenerate mass spectra where the above 
analysis is not applicable.}
\end{figure}
the correct baryon asymmetry and dark matter abundance can be generated 
only  if $\Gamma_3\gg\Gamma_2$, i.e. if the lepton asymmetry is generated 
by the same heavy neutrino which is responsible for the entropy production. 
The points of the parameters space where this is the case are schematically 
depicted in Fig.\ref{ParameterSpace}.

\subsection{Quasidegenerate mass spectrum}
For a quasidegenerate mass spectrum of the heavy neutrinos the \CP-violating parameter
is resonantly enhanced. In this regime the vertex contribution is much smaller than 
the self-energy one and can be neglected. The size of the \CP-violating parameter is 
controlled by the difference of the right-handed neutrino masses, see \eqn\eqref{epsilon_s}. 
If the mass difference is not too small, $R\gtrsim 10^3$, then one can neglect medium 
corrections to the masses \cite{Frossard:2012pc}. In this regime the $A_{ij}$ term 
in \eqref{epsilon_s} can still be neglected and we obtain 
\begin{align}
  \label{epsilon_s_approx}
  \epsilon^S_i \approx -\frac{\eta_{ij}}{R_{ij}} \approx 
 -\eta_{ij}\frac{M_i \Gamma_j}{\Delta M_{ji}^2}\,.
\end{align}
Since we enumerate the right-handed Majorana neutrinos such that 
$M_2/\sqrt{\Gamma_2}>M_3/\sqrt{\Gamma_3}$ and $M_2\sim M_3$ in the 
case under consideration then $\Gamma_3> \Gamma_2$. Combined with 
\eqref{epsilon_s_approx} this implies that $\epsilon_2>\epsilon_3$.
In other words, just like for a strongly hierarchical mass spectrum,
most of the lepton asymmetry is generated by the neutrino responsible 
for the entropy production. The required magnitude of the \CP-violating
parameter, $\epsilon \sim 10^{-7}$, can now be achieved even if $\eta_{ij}$  
is as small as $\sim 10^{-4}$. The condition \eqref{eq:2} ensures that 
$N_2$ decays before the sphaleron freezout so that the generated lepton  
asymmetry is converted into the baryon asymmetry. Therefore, successful 
leptogenesis is possible even if $M_2$ is as light as $\sim 10^4$ GeV. 

For smaller values of $R$ the enhancemet of the \CP-vi\-o\-la\-ting parameter 
is even stronger. However, one should keep in mind that for very small 
mass differences the usual quasiparticle approximation used in the present 
analysis breaks down \cite{DeSimone:2007pa}. Furthermore, the oscillating 
behavior of the heavy neutrino propagators leads to a suppression of the 
final asymmetry \cite{Garny:2011hg}. 

It is also important to note that since leptogenesis in this scenario can 
occur at temperatures much smaller than the ones where charge lepton Yukawa 
interactions enter in equilibrium, the flavour effects can become important
and affect the final result for the baryon asymmetry \cite{Beneke:2010dz}. 
This analysis however is beyond the scope of the present paper.

\section{Conclusions}

We have analyzed the generation of the baryon asymmetry of the Universe
in scenarios with an extended gauge sector where a keV scale sterile neutrino 
is a DM particle.  It was shown in \cite{Bezrukov:2009th} that 
the DM abundance can be controlled by the entropy produced in
the out of equilibrium decay of one of the other two heavier neutrinos.
Here we have found that the decays of the same neutrino can also lead
to the generation of a significant lepton asymmetry, leading to the
observed  baryon asymmetry of the Universe,  but at a price of a stronger 
bound on the extra gauge interaction scale and the heavy sterile neutrino 
masses.  For reasonable values of the model parameters the third
heavy neutrino does not contribute significantly to the 
asymmetry generation. For a hierarchical mass spectrum of the heavy 
neutrinos the resulting constraints on the masses of the model are
stronger than those in \cite{Bezrukov:2009th},
pushing the masses of the heavier sterile neutrinos and right handed
gauge bosons to the $\unit[10^{10}]{GeV}$ scale. For a quasidegenerate 
mass spectrum the \CP-violating parameter is resonantly enhanced and 
the required amount of the asymmetry can be produced even if the Majorana
neutrinos are as light as $\sim\unit[10^{4}]{GeV}$. 
 
Note that our scenario is incomplete in the sense that some
mechanism must explain the lightness of $N_1$ and the 
very tiny active-sterile mixings. Examples where these properties are
explained by flavour symmetries \cite{Lindner:2010wr,Barry:2011fp}
show that there might be a very interesting connection between dark 
matter, the baryon asymmetry of the Universe and neutrino flavour properties. 

\subsection*{Acknowledgements}
\noindent
The work of F.B. was partially supported by the Humboldt foundation. 
A.K. is supported by DFG under Grant KA-3274/1-1 ``Systematic analysis of 
baryogenesis in non-equilibrium quantum field theory''.



\end{document}